\documentclass{article}
\title{Analysis of constrained theories without use of primary constraints.}
\author{A.A. Deriglazov\footnote{alexei@ice.ufjf.br ~ On leave of
absence from Dept. Math. Phys., Tomsk Polytechnical University,
Tomsk, Russia.}}
\date{Dept. de Matematica, ICE, Universidade Federal de Juiz de Fora,\\
MG, Brasil.}
\begin{document}
\maketitle
\large

\begin{abstract}
It is shown that the Dirac approach to Hamiltonization of singular theories can be 
slightly modified in such a way that primary Dirac constraints do not appear in the process. 
According to the modified scheme, Hamiltonian formulation of singular theory is first order 
Lagrangian formulation, further rewritten in special coordinates. 
\end{abstract}

\noindent
%{\bf PAC codes:} \\
%{\bf Keywords:} Hamiltonian systems with constraints, Gauge theories

\section{Introduction}

Let us consider Lagrangian theory with the action and the equations of motion being
\begin{eqnarray}\label{1}
S=\int d\tau \bar L(q^A,\dot q^A), \qquad A=1,2,\cdots [A],
\end{eqnarray}
\begin{eqnarray}\label{2}
%\frac{\delta S}{\delta q^A}\equiv
\left(\frac{\partial \bar L}{\partial\dot q^A}\right)^{\bf{.}}-
\frac{\partial \bar L}{\partial q^A}=0.
\end{eqnarray}
It is supposed that the theory is singular  
\begin{eqnarray}\label{3}
rank\frac{\partial^2 \bar L}{\partial\dot q^A\partial\dot q^B}= [i]<[A]. 
\end{eqnarray}
According to Dirac [1], Hamiltonian formulation of the theory is obtained as follow. 
First step of Hamiltonization procedure is to define equations for   
the momenta $p_A$: $ ~ p_A=\frac{\partial\bar L}{\partial \dot q^A} ~$. The latter can be    
considered as algebraic equations for determining 
of velocities $\dot q^A$. According to the rank condition (\ref{3}), $[i]$ equations 
can be resolved for $\dot q^i$ and then substituted into the remaining ones. By construction, the 
resulting equations do not depend on $\dot q^A$ and are called primary constraints $\Phi_\alpha(q, p)$ 
of the Hamiltonian formulation. The equations $p_A=\frac{\partial\bar L}{\partial \dot q^A} ~$ 
are then equivalent to the following system  
\begin{eqnarray}\label{4}
\dot q^i=v^i(q^A, p_i, \dot q^\alpha),
\end{eqnarray}
\begin{eqnarray}\label{5}
\Phi_\alpha(q^A, p_\alpha, p_i)=0.
\end{eqnarray}
On second step one introduces an extended phase space with the coordinates $(q^A, p_A, v_\alpha)$. 
By definition, Hamiltonian formulation of the theory (\ref{2}) is the following system of 
equations on this space 
\begin{eqnarray}\label{6}
\dot q^A=\{q^A, H\}, ~  \dot p_A=\{p_A, H\}, \cr 
\Phi_\alpha(q^A, p_B)=0, \qquad
\end{eqnarray}
where $\{ ~ , ~ \}$ is the Poisson bracket, and it was denoted 
\begin{eqnarray}\label{7}
H(q^A, p_A, v^\alpha)=H_0(q^A, p_j)+v^\alpha\Phi_\alpha(q^A, p_B),
\end{eqnarray}
\begin{eqnarray}\label{8}
H_0=\left(p_i\dot q^i-\bar L+ \dot q^\alpha\frac{\partial\bar 
L}{\partial \dot q^\alpha}\right) 
\Biggr|_{\dot q^i\rightarrow v^i(q^A, p_j, \dot q^\alpha)}.
\end{eqnarray}
It is known [2] that the formulations (\ref{2}) and (\ref{6}) are equivalent. 

Aim of this part of the Dirac procedure is, in fact, to represent system of differential equations 
(\ref{2}) in the normal form (with derivatives of variables separated on the l.h.s. of equations). 
Adopting this point of view, we demonstrate below that Eq.(\ref{5}), as well as the 
variables $p_\alpha$, are not necessary in the process. In Sect. 2 we formulate 
Hamiltonization procedure for the theory (\ref{1})-(\ref{3}) without use of the 
momenta $p_\alpha$. As a consequence, the 
primary constraints (\ref{5}) do not appear in the formulation. 
Concerning to Eq.(\ref{4}), it has simple 
interpretation [3] as (invertible) change of variables on configuration-velocity 
space{\footnote{In this respect, the formulation presented below is more close to the case of 
nonsingular theory, where transition from the Lagrangian to the Hamiltonian formalism is 
change of variables.}}.
Expression  for the change will be obtained here as sufficient condition 
for conversion of equations of motion (see Eq.(\ref{17}) below) into the normal form.  
Equivalence with the Dirac procedure is shown in Sect.3. Some useful identities are collected 
in the Appendix.

\section{Hamiltonization without use of primary constraints}

We present here modification of standard Hamiltonization procedure [1, 2] which do not implies 
appearance of the primary constraints in the formalism.
 
{\bf{1 Step.}} System of differential equations of second order (\ref{2}) can be rewritten in a 
first order form. To achieve this, one introduces configuration-velocity space with the coordinates 
$(q^A, v^A)$. The initial system and the following one are equivalent{\footnote{The 
systems (\ref{2}) 
and (\ref{9}) are equivalent in the usual sense [2]: a) if $q_0(\tau), v_0(\tau)$ is solution 
of (\ref{9}), then $q_0(\tau)$ is solution of (\ref{2}); b) if $q_0(\tau)$ is solution 
of (\ref{2}) then $q_0(\tau), p_0(\tau)\equiv\dot q_0$ is solution of (\ref{9}).}}
\begin{eqnarray}\label{9}
\dot q^A=v^A, \qquad 
\bar M_{AB}\dot v^B=\bar K_A.
\end{eqnarray}
Here it was denoted
\begin{eqnarray}\label{11}
\bar M_{AB}(q, v)\equiv\frac{\partial^2 \bar L(q, v)}{\partial v^A\partial v^B}, \quad   
\bar K_A(q, v)\equiv\frac{\partial \bar L}{\partial q^A}-
\frac{\partial^2 \bar L}{\partial v^A\partial q^B}v^B.
\end{eqnarray}
We use symbols with bar to denote functions on the configuration-velocity space:
$\bar A\equiv\bar A(q^A, v^B)$, and symbols without bar for functions on 
extended phase space (see below)
\begin{eqnarray}\label{102}
A=A(q^A, p_j, v^\alpha)\equiv
\bar A(q^A, v^\alpha, v^i)|_{v^i(q^A, p_j, v^\alpha)}.
\end{eqnarray}
Without loss of generality [2], we suppose that the variables $q^A$ has been enumerated 
in a way adjusted with Eq.(\ref{3}), then 
\begin{eqnarray}\label{12}
\bar M_{AB}=\left(\begin{array}{cc}
\ \bar M_{ij} & \bar M_{i\beta}\\\bar M_{\alpha j} & \bar M_{\alpha\beta}
\end{array}\right), \qquad \qquad  \cr 
A=(i, \alpha), \quad i=1,2,\cdots [i], \quad \alpha =1,2,\cdots [\alpha]=[A]-[i], \cr 
rank \bar M_{AB}=[i], \quad det \bar M_{ij}\ne 0. \qquad 
\end{eqnarray}
Inverse matrix for $\bar M_{ij}$ is denoted 
$\tilde{\bar M}^{ij}$, one has $\bar M_{ij}\tilde {\bar M}^{jk}=\delta_i{}^k$.

{\bf{2 Step.}} According to the rank condition, $[i]$ equations among (\ref{9}) can be resolved 
in relation of $\dot v^i$:  
$\dot v^i=\tilde {\bar M}^{ij}(K_j-\bar M_{j\alpha}\dot v^\alpha)$
and then substituted into remaining $[\alpha]$ equations with the result being 
$[\bar M_{\alpha\beta}-\bar M_{\alpha i}\tilde{\bar M}^{ij}\bar M_{j\beta}]\dot v^\beta=
\bar K_\alpha-\bar M_{\alpha i}\tilde {\bar M}^{ij}\bar K_j$. It must be 
$rank[\bar M_{\alpha\beta}-\bar M_{\alpha i}\tilde{\bar M}^{ij}\bar M_{j\beta}]=0$ (if not, one would be able to resolve 
the equations in relation to some of $\dot v^\alpha$, in contradiction with (\ref{3})). Then one has 
the identity   
\begin{eqnarray}\label{13}
\bar M_{\alpha\beta}-\bar M_{\alpha i}\tilde {\bar M}^{ij}\bar M_{j\beta}=0.
\end{eqnarray}
As a consequence, the following quantities 
\begin{eqnarray}\label{14}
\bar C_\alpha{}^A\equiv
\left(-\bar M_{\alpha j}\tilde {\bar M}^{ji},  \delta_\alpha{}^\beta\right),
\end{eqnarray}
form a basis on space of null-vectors of the matrix $\bar M_{AB}$ (see also [3]) 
\begin{eqnarray}\label{15}
\bar C_\alpha{}^A\bar M_{AB}=0.
\end{eqnarray}
In the result, our equations (\ref{9}) acquire the equivalent form
\begin{eqnarray}\label{16}
\dot q^A=v^A, 
\end{eqnarray}
\begin{eqnarray}\label{17}
\dot v^i=\tilde {\bar M}^{ij}(\bar K_j-\bar M_{j\alpha}\dot v^\alpha,)
\end{eqnarray}
\begin{eqnarray}\label{18}
\bar K_\alpha-\bar M_{\alpha i}\tilde {\bar M}^{ij}\bar K_j\equiv \bar C_\alpha{}^A\bar K_A=0,
\end{eqnarray}
where the system (\ref{18}) is an algebraic one, (also, it may contain functionally 
dependent equations).

{\bf{3 Step.}} Eq.(\ref{17}) is not in the normal form. Then one asks on existence of special 
coordinates $(\tilde q^A, \tilde v^A)$ on $(q^A, v^A)$-space which reduce
the system to the normal form.  We demonstrate here the following statement.

For singular theory (\ref{1})-(\ref{3}) there is exist invertible change of variables 
of the form
\begin{eqnarray}\label{19}
\left(
\begin{array}{c}
q^A \\
v^\alpha \\
v^i
\end{array}
\right)
\leftrightarrow
\left(
\begin{array}{c}
q^A \\
v^\alpha \\
p_i
\end{array}
\right), \quad 
v^i=v^i(q^A, p_j, v^\alpha), \quad 
rank\frac{\partial v^i}{\partial p_j}=[i],
\end{eqnarray}
which reduces the system (\ref{17}) to the normal form. 

To make a proof, let us write Eq.(\ref{17}) in terms of new variables (see our notations 
(\ref{102}))
\begin{eqnarray}\label{20}
M_{ij}\frac{\partial v^i}{\partial p_k}\dot p_k+
\left(M_{ij}\frac{\partial v^j}{\partial v^\alpha}+M_{i\alpha}\right)\dot v^\alpha=
K_i-M_{ij}\frac{\partial v^j}{\partial q^A}v^A.
\end{eqnarray}
It reduces to the normal form if the functions $v^i$ obey the equations 
(note that $M_{ij}$ is the only invertible matrix at hand)
\begin{eqnarray}\label{21}
\frac{\partial v^i}{\partial p_j}=\tilde M^{ij}, \qquad
\frac{\partial v^i}{\partial v^\alpha}=-\tilde M^{ij}M_{j\alpha}.
\end{eqnarray}
These equations determine $v^i$ modulo of an arbitrary function $A(q^A)$, the latter will be 
omitted in what follow. Suppose that there is exist a solution of Eq.(\ref{21}),  
and let $p_i(q^A, v^A)$ be inverse transformation. Expression for the latter can be easily 
obtained. Actually, using Eq.(\ref{21}), derivatives of the identity 
$v^i(q^A, p_i(q^A, v^A), v^\alpha)\equiv v^i$ 
can be written in the form 
$\frac{\partial p_i}{\partial v^j}=\bar M_{ij}=
\frac{\partial^2 \bar L}{\partial v^i\partial v^j}$, ~ 
$\frac{\partial p_i}{\partial v^\alpha}=\bar M_{i\alpha}=
\frac{\partial^2 \bar L}{\partial v^\alpha\partial v^i}$.
Solution of the system gives expression for $v^i$ in implicit form
\begin{eqnarray}\label{22}
p_i=\frac{\partial\bar L(q^A, v^i, v^\alpha)}{\partial v^i}.
\end{eqnarray}
It is necessary condition which turns out to be a sufficient condition also. Actually, 
Eq.(\ref{22}) is invertible due to Eq.(\ref{3}), and defines some function 
$v^i(q^A, p_j, v^\alpha)$. Then derivatives of the identity 
$p_i\equiv\frac{\partial\bar L}{\partial v^i}|_{v^i}$ 
imply immediately the equations (\ref{21}) (see also Appendix).

Curiously enough, the reasoning works for non singular theory also. To find normal form for 
non singular equations $\bar M_{AB}\dot v^B=\bar K_A$, one applies the inverse matrix 
$\tilde{\bar M}_{AB}$. Equivalently, one can search for a change $v^A(q^A, p_A)$ which reduces 
the system to the normal form. It gives the standard Hamiltonian formulation of the theory. 

{\bf{4 Step.}} Let us rewrite the equations (\ref{16})-(\ref{18}) in the coordinates (\ref{19}). 
Properties of the transformation $v^i=v^i(q^A, p_j, v^\alpha)$ are presented in the Appendix. 
In particular, it turns out to be linear on $v^\alpha$, and has the representation 
\begin{eqnarray}\label{23}
v^i(q^A, p_j, v^\alpha)=\frac{\partial H_R}{\partial p_i},
\end{eqnarray}
where 
\begin{eqnarray}\label{24}
H_R\equiv H_0(q^A, p_j)-f_\alpha(q^A, p_j)v^\alpha, \qquad  
f_\alpha\equiv\frac{\partial\bar L}{\partial v^\alpha}\Biggr|_{v^i},
\end{eqnarray}
and $H_0$ coincides with the Hamiltonian (\ref{8}) of the Dirac scheme. Moreover, in  
the coordinates (\ref{19}) all the quantities of the first order Lagrangian formulation can 
be rewritten in terms of the functions $H_0$ and $f_\alpha$, see Appendix. Using 
Eqs.(\ref{15}), (\ref{23}), (\ref{24}), 
(\ref{201}), (\ref{207}), the system (\ref{16})-(\ref{18}) can be identically rewritten in the form 
\begin{eqnarray}\label{25}
\dot q^\alpha=v^\alpha, 
\end{eqnarray}
\begin{eqnarray}\label{26}
\dot q^i=\frac{\partial H_R}{\partial p_i}, \qquad \qquad 
\dot p_i=-\frac{\partial H_R}{\partial q^i}, 
\end{eqnarray}
\begin{eqnarray}\label{27}
\bigtriangleup_{\alpha\beta}(q^A, p_j)v^\beta+ H_\alpha(q^A, p_j)=0. 
\end{eqnarray}
Here it was denoted
\begin{eqnarray}\label{28}
\bigtriangleup_{\alpha\beta}=\{f_\alpha, f_\beta\}_R-
\left(\frac{\partial f_\alpha}{\partial q^\beta}-\frac{\partial f_\beta}{\partial q^\alpha}\right), \quad
H_\alpha=-\{f_\alpha, H_0\}_R-
\frac{\partial H_0}{\partial q^\alpha}, 
\end{eqnarray}
and $\{ ~ , ~ \}_R$ is reduced Poisson bracket (i.e. the Poisson bracket on $q^i, p_i$- 
subspace). For any functions $A, B$ of the variables $(q^A, p_i, v^\alpha)$ it is defined by 
\begin{eqnarray}\label{29}
\{A,~ B\}_R=\frac{\partial A}{\partial q^i}\frac{\partial B}{\partial p_i}-
\frac{\partial A}{\partial p_i}\frac{\partial B}{\partial q^i}.
\end{eqnarray}
It finishes first stage of the Hamiltonization procedure. Differential equations  
are presented in the normal form (\ref{26}). One notes also that algebraic equations (\ref{27}) 
turn out to be linear on $v^\alpha$. 
They coincide with the Dirac second stage system (for revealing of secondary constraints): 
$\{\Phi_\alpha, H\}=0$ (see also the next section).

In resume, initial equations of motion (\ref{2}) for singular theory have been rewritten in 
special coordinates (\ref{19}) on the configuration-velocity space. 
Motivation for making of  
the change was to find normal form of differential equations (\ref{17}).
It turns out to be 
sufficient to transform the variables $v^i$ only, defining the new variables $p_i(v^j)$ in 
accordance with Eq.({\ref{22}). 
One concludes that representation of singular Lagrangian 
dynamics in the normal form (\ref{25})-(\ref{27}) do not implies any extension of the 
configuration-velocity space. In particular, the primary Dirac constraints do not appear in 
the formulation. We have demonstrated also that the Dirac second stage system 
$\{\Phi_\alpha, H\}=0$ represents part of Lagrangian equations of motion 
(\ref{18}), rewritten in the coordinates $(q^A, p_i, v^\alpha)$.
By construction, the systems (\ref{2}) and (\ref{25})-(\ref{27}) are equivalent.
It would be interesting to analyze local symmetries [4, 3] in this framework, as well as to compare 
the reduced scheme with other modifications of the Dirac procedure [5, 6]. 
   
\section{Equivalence with the Dirac procedure}

We discuss here relation among the system (\ref{25})-(\ref{27}) and the standard Hamiltonian 
formulation (\ref{6}). Starting from (\ref{25})-(\ref{27}), let us define an equivalent system on 
extended space with the coordinates $(q^A, p_j, v^\alpha, p_\alpha)$. By definition,  
$(q^A, p_j, v^\alpha)$-sector obeys the Eqs.(\ref{25})-(\ref{27}), while for 
$p_\alpha$ one writes the primary Dirac constraints  
\begin{eqnarray}\label{30}
\Phi_\alpha\equiv p_\alpha-f_\alpha(q^A, p_j)=0. 
\end{eqnarray} 
Then Eq.(\ref{27}) can be rewritten identically in the form $\{\Phi_\alpha, H\}=0$, while
the following equation: 
\begin{eqnarray}\label{31}
\dot p_\alpha=-\frac{\partial H}{\partial q^\alpha} 
\end{eqnarray}
is a consequence of 
Eqs.(\ref{25})-(\ref{27}), (\ref{30}), and thus can be added to this system. 
After that, Eq.(\ref{27}) 
turns out to be consequence of others, and can be omitted. The resulting system (\ref{25}), 
(\ref{26}), (\ref{30}), (\ref{31}) coincides exactly with the Hamiltonian formulation (\ref{6})
(modulo of notations).

This discussion reveals that the only role, playing by $p_\alpha$ in the Dirac formulation, is 
to represent equations of motion in completely symmetric form, with the Poisson bracket defined in 
relation to all variables $q^A, p_A$. The momenta $p_\alpha$ are, in fact, "Hamiltonian ghosts" of 
singular theory. The symmetrization seems to be unnecessary ingredient of this stage. 
Further stages of the Dirac procedure (on revealing of higher-stage 
constraints) do not require the symmetric form of equations.
It may be more convenient to make further steps by using of the reduced formulation, 
and then to symmetrize the resulting equations.

\section{Conclusion}

In this work 
it was demonstrated that Hamiltonian formulation for singular theory (\ref{1})-(\ref{3}) can be 
considered as the first order Lagrangian formulation (\ref{9}) on the space $(q^A, v^i, v^\alpha)$,  
further rewritten in special coordinates $(q^A, p_i, v^\alpha)$ of this space.
The corresponding change of variables was find from the requirement 
that equations (\ref{9}) acquire normal form in the new coordinate system. 
The transformation 
$v^i(q^A, p_j, v^\alpha)$, defined implicitly by Eq.(\ref{22}), turns out to be sufficient condition 
to this end. Search for the normal form do not implies use of an additional variables, 
as well as the primary Dirac constraints. The resulting formulation (\ref{25})-(\ref{27}) involves 
algebraic equations (\ref{27}) linear on $v^\alpha$. They coincide with the second stage 
equations of the Dirac procedure: $\{\Phi_\alpha, H\}=0$. The system (\ref{25})-(\ref{27}) and 
the standard one (\ref{6}) are equivalent.

For the scheme presented one writes, schematically
\begin{eqnarray}\label{32}
q^A\rightarrow (q^A, v^i, v^\alpha)\leftrightarrow (q^A, p_i, v^\alpha). 
\end{eqnarray}
It can be compared with the standard approach: according to [2], Hamiltonian formalism involves more  
variables
\begin{eqnarray}\label{33}
q^A\rightarrow (q^A, v^i, v^\alpha)\rightarrow (q^A, v^i, v^\alpha, p_i, p_\alpha). 
\end{eqnarray}
Dynamics of $(q^A, v^A)$-sector is governed by 
the Lagrangian first order equations (\ref{9}), while 
$p_A=\frac{\partial\bar L}{\partial v^A}$.

\section{Acknowledgments}
Author would like to thank the Brazilian foundations CNPq and FAPEMIG 
for financial support.

\section{Appendix}

We present here properties of the function $v^i(q^A, p_j, v^\alpha)$ defined implicitly by Eq.(\ref{22}), 
as well as properties of the Lagrangian $L$ in terms of the coordinates (\ref{19}): 
$L(q^A, p_j, v^\alpha)\equiv\bar L(q^A, v^i, v^\alpha)|_{v^i}$. In particular, both functions turn 
out to be at most linear on $v^\alpha$ in any singular theory (\ref{1})-(\ref{3}).

Derivatives of the identity 
$p_i\equiv\frac{\partial\bar L(q^A, v^i, v^\alpha)}{\partial v^i}|_{v^i(q^A, p_j, v^\alpha)}$ 
give the relations (see our notations (\ref{102}), (\ref{12}))
\begin{eqnarray}\label{201}
\frac{\partial v^i}{\partial p_j}=\tilde M^{ij}, \qquad
\frac{\partial v^i}{\partial v^\alpha}=-\tilde M^{ij}M_{j\alpha}, \qquad
\frac{\partial v^i}{\partial q^A}=-\tilde M^{ij}
\frac{\partial^2 \bar L}{\partial v^j\partial q^A}\Biggr|_{v^i}.
\end{eqnarray}
Then the identity (\ref{13}) acquires the form 
$\frac{\partial}{\partial v^\alpha}\left(\frac{\partial\bar L}{\partial v^\beta}|_{v^i}\right)=0$, 
so the quantity 
\begin{eqnarray}\label{202}
f_\alpha(q^A, p_j)\equiv\frac{\partial\bar L}{\partial v^\alpha}\Biggr|_{v^i}.
\end{eqnarray}
does not depend on $v^\alpha$.
One finds $\frac{\partial f_\alpha(q^A, p_j)}{\partial p_i}=\tilde M^{ij}M_{j\alpha}$, so the 
quantity $\tilde M^{ij}M_{j\alpha}$ does not depend on $v^\alpha$ also. Moreover, Eq.(\ref{202}) can 
be identically rewritten in terms of $L$
\begin{eqnarray}\label{203}
f_\alpha(q^A, p_j)=-\frac{\partial}{\partial v^\alpha}
\left(p_iv^i-L\right), 
\end{eqnarray}
which implies that the quantity     
\begin{eqnarray}\label{204}
H_R(q^A, p_j, v^\alpha)\equiv (p_iv^i-L),
\end{eqnarray}
is at most linear on $v^\alpha$. Integral of Eq.(\ref{203}) gives the following expression for $H_R$
\begin{eqnarray}\label{205}
H_R(q^A, p_j, v^\alpha)=H_0(q^A, p_j)-v^\alpha f_\alpha(q^A, p_j),
\end{eqnarray}
where $H_0$ does not depend on $v^\alpha$.
Expression for this quantity in terms of initial Lagrangian can be obtained by comparison of 
Eqs.(\ref{204}), (\ref{205})
\begin{eqnarray}\label{206}
H_0(q^A, p_j)=\left(\frac{\partial\bar L}{\partial v^A}v^A-\bar L\right)\Biggr|_{v^i}.
\end{eqnarray}
Note that it coincides exactly with the Hamiltonian $H_0$ (\ref{8}) of the 
standard scheme.
From Eq.(\ref{204}) one obtains useful relations 
\begin{eqnarray}\label{207}
\frac{\partial H_R}{\partial p_i}=v^i, \qquad
\frac{\partial H_R}{\partial q^A}=-\frac{\partial\bar L}{\partial q^A}\Biggr|_{v^i}, \qquad
\frac{\partial H_R}{\partial v^\alpha}=-f_\alpha(q^A, p_j).
\end{eqnarray}
In particular, the first equation states that our function $v^i$ is at most linear on $v^\alpha$ 
and has the representation
\begin{eqnarray}\label{208}
v^i(q^A, p_j, v^\alpha)=\frac{\partial H_0}{\partial p_i}-\frac{\partial f_\alpha}{\partial p_i}v^\alpha.
\end{eqnarray}
It implies (see Eq.(\ref{201})) that the matrix $\tilde M^{ij}$ is at most linear on $v^\alpha$, and has the 
representation $\tilde M^{ij}=\frac{\partial^2 H_R}{\partial p_i\partial p_j}$.
At last, Eqs.(\ref{204}), (\ref{205}), (\ref{208}) imply that $L$ is at most linear on 
$v^\alpha$ and has the representation
\begin{eqnarray}\label{207}
L(q^A, p_j, v^\alpha)=p_i\frac{\partial H_R}{\partial p_i}-H_R.
\end{eqnarray}
In resume, we have demonstrated, in fact, that in the coordinates $(q^a, p_i, v^\alpha)$ all  
quantities of the first order Lagrangian formulation (\ref{9})-(\ref{11}) can be rewritten in terms of 
the functions $H_0(q^A, p_j), ~ f_\alpha(q^A, p_j)$.

\end{document}